\documentclass[twocolumn,twocolappendix]{aastex631}

\usepackage{amsmath}
\usepackage{graphicx}

\received{2023 December 31}
\revised{2024 March 27}
\accepted{2024 April 4}

\shorttitle{PAC VI. QSO.}
\shortauthors{Gui et al.}

\begin{document}

\title{Photometric Objects Around Cosmic Webs (PAC). VI. High Satellite Fraction of Quasars}

\author[0000-0002-0359-4284]{Shanquan Gui}
\affiliation{Department of Astronomy, School of Physics and Astronomy, Shanghai Jiao Tong University, Shanghai, 200240, People’s Republic of China}

\author[0000-0002-7697-3306]{Kun Xu}
\affil{Department of Astronomy, School of Physics and Astronomy, Shanghai Jiao Tong University, Shanghai, 200240, People’s Republic of China}
\affil{Institute for Computational Cosmology, Department of Physics, Durham University, South Road, Durham DH1 3LE, UK}

\author[0000-0002-4534-3125]{Y.P. Jing}
\affil{Department of Astronomy, School of Physics and Astronomy, Shanghai Jiao Tong University, Shanghai, 200240, People’s Republic of China}
\affil{Tsung-Dao Lee Institute, and Shanghai Key Laboratory for Particle Physics and Cosmology, Shanghai Jiao Tong University, Shanghai, 200240, People’s Republic of China}

\author{Donghai Zhao}
\affil{Key Laboratory for Research in Galaxies and Cosmology, Shanghai Astronomical Observatory, Shanghai, 200030, People's Republic of China}
\affil{Department of Astronomy, School of Physics and Astronomy, Shanghai Jiao Tong University, Shanghai, 200240, People’s Republic of China}

\author{Hongyu Gao}
\affil{Department of Astronomy, School of Physics and Astronomy, Shanghai Jiao Tong University, Shanghai, 200240, People’s Republic of China}

\correspondingauthor{Y.P. Jing, Donghai Zhao}
\email{ypjing@sjtu.edu.cn, dhzhao@shao.ac.cn}



\begin{abstract}

The Photometric objects Around Cosmic webs (PAC) approach developed in Xu et al. (2022b) has the advantage of making full use of spectroscopic and deeper photometric surveys. With the merits of PAC, the excess surface density $\bar{n}_2w_{{\rm{p}}}$ of neighboring galaxies can be measured down to stellar mass $10^{10.80}\,M_{\odot}$ around quasars at redshift $0.8<z_{\rm{s}}<1.0$, with the data from the Sloan Digital Sky Survey IV (SDSS-IV) extended Baryon Oscillation Spectroscopic Survey (eBOSS) and the Dark Energy Spectroscopic Instrument (DESI) Legacy Imaging Surveys. We find that $\bar{n}_2w_{{\rm{p}}}$ generally increases quite steeply with the decrease of the separation.  Using subhalo abundance matching method, we can accurately model the $\bar{n}_2w_{{\rm{p}}}$ both on small and large scales. We show that the steep increase of the $\bar{n}_2w_{{\rm{p}}}$ towards the quasars requires that a large fraction $f_{\mathrm{sate}}=0.29_{-0.06}^{+0.05}$ of quasars should be satellites in massive halos, and find that this fraction measurement is insensitive to the assumptions of our modeling. This high satellite fraction indicates that the subhalos have nearly the same probability to host quasars as the halos for the same (infall) halo mass, and the large scale environment has negligible effect on the quasar activity. We show that even with this high satellite fraction, each massive halo on average does not host more than one satellite quasar due to the sparsity of quasars.

\end{abstract}



\section{Introduction} \label{sec:intro}

The large-scale structure of the universe is primarily influenced by dark matter that only undergoes gravitational interactions. 
Various luminous tracers, such as luminous red galaxies (LRGs), quasars, and emission line galaxies, are employed to understand the distribution of the underlying dark matter. Quasars are the most luminous tracer among them and can reach the high redshift ($z_{\rm{s}}\footnote{Throughout the paper, we use $z_{\rm{s}}$ for spectroscopic redshift and $z$ for the $z$-band magnitude.} > 2.0$), offering invaluable information for investigating structure formation and cosmology in the early universe \citep{2020MNRAS.499..210N,2020ApJ...901..153D,2021MNRAS.500.1201H,2022JCAP...08..024B}. Moreover, quasars are believed to be powered by the mass accretion onto supermassive black holes (SMBHs) located at the center of each galaxy \citep{1982MNRAS.200..115S}. The strong feedback from the SMBHs regulates the formation of their host galaxies, shaping a process of co-evolution \citep{2014ARA&A..52..589H}. This underscores the importance of understanding quasars in the broader context of galaxy formation.  

In the realm of both cosmological and galaxy formation studies employing quasars, it is crucial to accurately delineate the connection between quasars and the underlying dark matter, which often involves precise measurements of the number density and multi-scale clustering of quasars. Since the pioneering work of \cite{1981ApJ...247..762O}, extensive efforts have been dedicated to quantifying quasar clustering through correlation functions. Among these, the two-point correlation function (2PCF) stands out as the predominant method. The availability of datasets such as the 2dF Quasi-Stellar Object Redshift Survey (2QZ) \citep{2004MNRAS.349.1397C}, SDSS-I/II/III/BOSS, and SDSS-IV/eBOSS \citep{2002AJ....123..567S,2010AJ....139.2360S,2011AJ....142...72E,2020ApJS..250....8L}, along with the emergence of surveys like the Dark Energy Spectroscopic Instrument (DESI) \citep{2016arXiv161100036D}, Subaru Prime Focus Spectrograph (PFS) survey \citep{2018SPIE10702E..1CT}, and Euclid \citep{2011arXiv1110.3193L}, has enabled high-precision measurements of the 2PCF at large scales ($>1\,h^{-1}\rm{Mpc}$).
However, accurately measuring quasar clustering at small scales ($<1\,h^{-1}\rm{Mpc}$) remains challenging due to the technical issue of fiber collisions and the low quasar number density. While several up-weighting schemes have been proposed to address the fiber collision problem \citep{1998ApJ...494....1J,2012ApJ...756..127G,2014MNRAS.444..476R,2017MNRAS.467.1940H,2017MNRAS.472L..40P,2019ApJ...872...26Y,2020JCAP...06..057S}, and additional observations have been conducted to increase the number of quasar pairs selected from 2QZ and SDSS images \citep{2006AJ....131....1H,2008ApJ...678..635M,2012ApJ...755...30R,2017MNRAS.468...77E}, the inherent difficulties persist. Moreover, numerous studies seek to enhance small-scale measurements by cross-correlating quasars with other spectroscopic tracers, such as LRGs \citep{2002MNRAS.332..529K,2009MNRAS.397.1862P,2011ApJ...726...83M,2013ApJ...773..175Z,2013ApJ...778...98S}. Although small-scale measurements can be more precise than relying solely on quasar auto-correlation, the obstacles remain as the the other tracers are also sparse at high redshift and the fiber collision issue may persist.

On the modeling front, various studies aim to establish connections between observed quasars and the underlying dark matter. In the simplest scenario, the linear bias factor can be determined by examining the relative amplitude of dark matter and quasars in the 2PCF at large scales \citep{1998ApJ...503L...9J}. This linear bias factor, in turn, can be employed to provide a rough estimate for the typical mass of the dark matter halos that host quasars \citep{2009ApJ...697.1634R}. The halo occupation distribution (HOD) method has been utilized to model the 2PCF for an extensive range. Numerous studies have consistently found that the typical mass of quasars falls within the range of $10^{12}\, h^{-1}M_{\odot} \sim 10^{13}\,h^{-1} M_{\odot}$, showing a remarkable degree of independence with respect to redshift or quasar luminosity \citep{2004MNRAS.355.1010P,2005MNRAS.356..415C,2006ApJ...638..622M,2007ApJ...654..115C,2007ApJ...658...85M,2007AJ....133.2222S,2008MNRAS.383..565D,2009MNRAS.397.1862P,2012ApJ...755...30R,2012MNRAS.424..933W,2013ApJ...778...98S,2018MNRAS.477...45M,2020ApJ...891...41P}.

The satellite fraction of quasars can be derived using the HOD framework, and this measurement proves particularly sensitive to small scale clustering. As noted previously, accurately measuring the small-scale clustering of quasars poses a challenge, leading to discrepant results in the literature \citep{2011ApJ...741...15S,2012MNRAS.424.1363K,2015MNRAS.446.1874L,2019MNRAS.486..274E,2019MNRAS.487..275G,2021MNRAS.504..857A}. For instance, the satellite fraction of quasars is determined as $0.068_{-0.023}^{+0.034}$ and $0.099_{-0.036}^{+0.046}$ through the cross-correlation between quasars and LRGs under two distinct quasar HOD models at redshift $0.3 - 0.9$ \citep{2013ApJ...778...98S}. Additionally, it is measured to be $7.3_{-1.5}^{+0.6}\times 10^{-4}$ using the auto-correlation of quasars at redshift $0.4 - 2.5$ \citep{2012ApJ...755...30R}. However, it is important to determine this fraction, as the information is crucial not only for determining the small scale clustering of quasars (e.g., the close pairs) but also for studying the halo environment effect on the formation of quasars.

In this study, to achieve precise measurements of quasar clustering across various scales and accurately constrain the quasar-halo connection and satellite fraction, we utilize the Photometric Objects Around Cosmic Webs (PAC) approach. On the basis of \citet{2011ApJ...734...88W}, \citet{2022ApJ...925...31X} developed the PAC method, which capitalizes on the benefits of leveraging both spectroscopic and deeper photometric surveys. PAC can measure the excess surface density $\bar{n}_2w_{\rm{p}}$ of photometric objects with specific physical properties around spectroscopic tracers. With the PAC method, \citet{2022ApJ...939..104X} and \citet{2023ApJ...944..200X} precisely determined the stellar-halo mass relation (SHMR) and galaxy stellar mass function (GSMF) down to the stellar mass limit where the spectroscopic sample is already highly incomplete, emphasizing the effectiveness of the PAC method. Given the absence of fiber collision problems between photometric and spectroscopic surveys, PAC becomes a valuable tool for accurately measuring the environment on $h^{-1} \rm{Mpc} $ scale for quasars. Additionally, PAC enables the assessment of the spatial distribution of galaxies with diverse properties around quasars, offering extensive information to study the quasar-halo connection. We use the quasar sample from the Sixteenth Data Release (DR16) in SDSS-IV/eBOSS \citep{2016ApJS..224...34P,2020ApJS..249....3A,2020ApJS..250....8L} and the photometric sample from the Ninth Data release (DR9) in DESI Legacy Imaging Surveys \citep[Legacy Surveys,][]{2019AJ....157..168D}. We concentrate on a limited redshift range of $0.8 < z_{\rm{s}} < 1.0$ due to constraints posed by the survey depth of the Legacy Surveys. We employ the subhalo abundance matching method (SHAM) to model the $\bar{n}_{2}w_{\rm{p}}$ measurements from PAC and simultaneously constrain the SHMR for galaxies and the quasar-halo connection. Our analysis of the quasar-halo connection reveals a substantial satellite fraction $f_{\mathrm{sate}}=0.29_{-0.06}^{+0.05}$. This satellite fraction implies that a subhalo has nearly the same chance to host a quasar as a halo for the same (infall) mass, and quasars are more likely triggered by the galactic internal process instead of the large scale (halo) environment.


The paper is structured as follows: Section \ref{sec:obs} provides a description of PAC and its measurements. Section \ref{sec:sim} covers the N-body simulation, the subhalo abundance matching method description, and the results of modeling, along with corresponding discussions. Finally, Section \ref{sec:dis} presents our conclusions. Throughout the paper, we adopt a spatially flat $\Lambda \mathrm{CDM}$ cosmology  with $\Omega_{\mathrm{m},0} = 0.268$, $\Omega_{\Lambda,0} = 0.732$, and $H_0 = 100\,h\mathrm{km s^{-1}Mpc^{-1}}=71 \,\mathrm{kms^{-1}Mpc^{-1}}$\citep{2013ApJS..208...19H}.

\section{OBSERVATIONS AND MEASUREMENTS} \label{sec:obs}

In this section, we provide a brief overview of the PAC concept, and detail the spectroscopic and photometric samples utilized in this study. We assess the completeness of the photometric sample in terms of stellar mass. Subsequently, we present the measurement of $\bar{n}_2w_{\rm{p}}$.

\subsection{Photometric objects Around Cosmic webs (PAC)} \label{sec:obs:pac}

Suppose we have two populations of objects, one from a spectroscopic catalog and the other from a photometric catalog. PAC can measure the excess surface density $\bar{n}_2w_{\rm{p}}$ of photometric objects with certain physical properties around spectroscopic objects across a wide range of scales:
\begin{equation}
    \bar{n}_2w_{\rm{p}}(r_{\rm{p}}) = \frac{\bar{S}_2}{r_1^2}\omega_{12,\rm{weight}}(\theta)\,\,,\label{eaq:1}
\end{equation}
where $\bar{n}_2$ is the mean number density of photometric objects and $\bar{S}_2$ is the mean angular surface density of photometric objects. $r_1$ is the comoving distance to the spectroscopic objects. $w_{\rm{p}}(r_{\rm{p}})$ and $\omega_{12,\rm{weight}}(\theta)$ are the projected cross-correlation function (PCCF) and the weighted angular cross-correlation function (ACCF) between spectroscopic objects and photometric objects, with $r_{\rm{p}}=r_1\theta$. To account for the case that $r_{\rm{p}}$ varies with $r_1$ at fixed $\theta$ caused by the redshift distribution of the spectroscopic objects, $\omega_{12}(\theta)$ is weighted by $1/r^2_1$. With PAC, we can take advantage of the deep photometric surveys to statistically obtain its rest-frame physical properties without making use of photo-z.
We list the key steps of PAC as follows and refer to \cite{2022ApJ...925...31X} for a detailed aacount:
\begin{enumerate}
    \item Divide spectroscopic objects into several sub-samples with narrower redshift bins to limit the range of $r_1$ in each bin.
    \item Assign the median redshift in each redshift bin to the entire photometric sample. The physical properties of photometric objects can be estimated through spectral energy distribution (SED) fitting with the assigned redshift. Consequently, in each redshift bin, there is a physical property catalog for the photometric sample.
    \item In each redshift bin, select photometric objects with specific physical properties and calculate $\bar{n}_2w_{\rm{p}}(r_{\rm{p}})$ according to Equation \ref{eaq:1}. Foreground and background objects with incorrect redshifts are effectively eliminated through ACCF, leaving only the photometric objects around spectroscopic objects with correct physical properties.
    \item Combine the results from different redshift bins by averaging with appropriate weights.
\end{enumerate}

\subsection{Spectroscopic and Photometric Samples} \label{sec:obs:sample}
For the photometric sample, we utilize the DR9 catalog from the Legacy Imaging Surveys \citep{2019AJ....157..168D}, which includes the Dark Energy Camera Legacy Survey (DECaLS), the Mayall $z$-band Legacy Survey (MzLS), and the Beijing-Arizona Sky Survey (BASS)\footnote{\url{https://www.legacysurvey.org/dr9/catalogs/}}. They cover approximately $14000\, {\rm{deg}}^2$ in the Northern and Southern Galactic caps (NGC and SGC) with $9000\, {\rm{deg}}^2$ at $\rm{decl.}\leq32^{\circ}$ and $5100\, {\rm{deg}}^2$ at $\rm{decl.} > 32^{\circ}$. The sources are extracted using \texttt{Tractor} \citep{2016AJ....151...36L} and subsequently modeled with profiles convolved with a specific point spread function (PSF). These profiles include a delta function for point sources, and exponential law, de Vaucouleurs law, and a S\'ersic profile for extended sources. We use their best-fit model magnitudes throughout the paper with the Galactic extinction corrected with the maps of \citet{1998ApJ...500..525S}. We merely include the footprints observed at least once in $g$, $r$ and $z$ bands. To remove bright stars and bad pixels, we use MASKBITS \footnote{\url{https://www.legacysurvey.org/dr9/bitmasks/}} offered by the Legacy Surveys. 
\begin{figure*}
	\centering
	\includegraphics[scale=0.22]{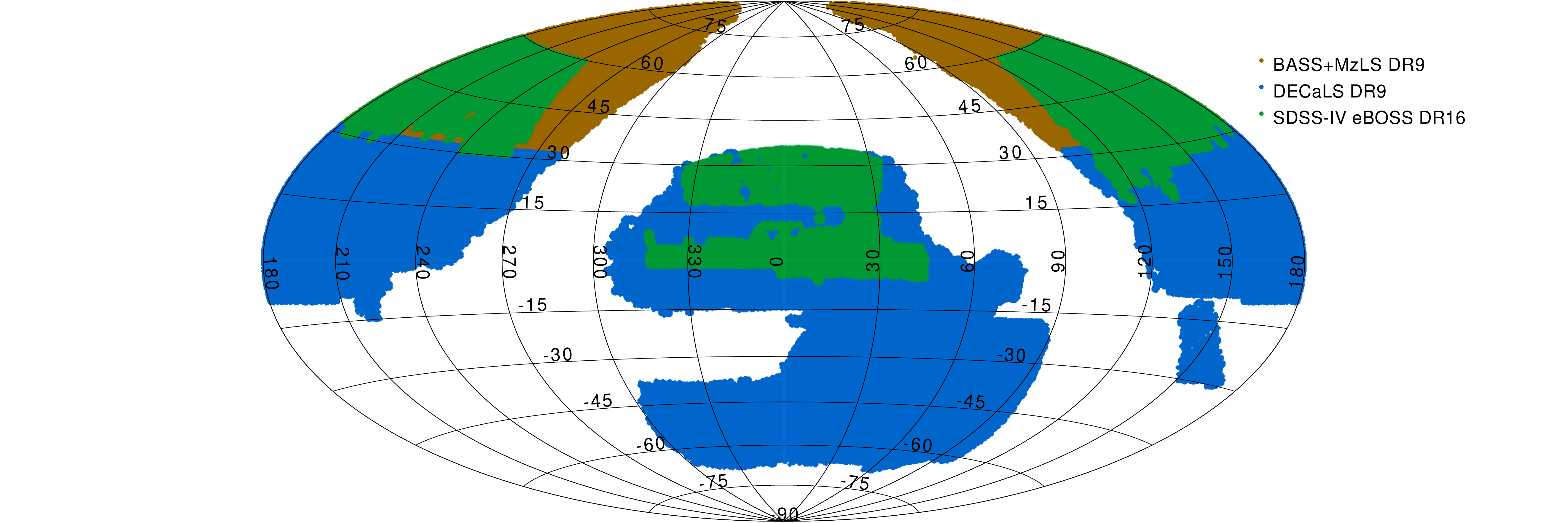}
	\caption{Spatial coverage of the two samples used in this work. The photometric sample is the Legacy Imaging Surveys DR9 where the DECaLS part is shown in blue color and the BASS+MzLS part is represented by dark orange color. The spectroscopic sample is SDSS-IV eBOSS DR16 quasars shown in blue color.}
	\label{f01}
\end{figure*}

For the spectroscopic sample, we utilize the SDSS-IV eBOSS DR16 quasar sample\footnote{\url{https://data.sdss.org/sas/dr16/eboss/lss/catalogs/DR16/}} which does not include SDSS-I/II/III quasars, within the redshift range $0.8<z_{\rm{s}}<1.0$. The sample comprises of 32,291 quasars. This redshift range is selected based on the depth of the Legacy surveys. The majority of eBOSS quasars were targeted by a {\texttt{CORE}} algorithm, which was applied to SDSS {\texttt{SURVEY-PRIMARY} point sources with $g<22$ or $r<22$ \textbf{\citep{2015ApJS..221...27M}}. These point sources were further selected by the XDQSOz algorithm \citep{2012ApJ...749...41B}, which imposed a probability higher than $20\%$ of being a quasar at redshift $z_{\rm{s}}>0.9$, with an additional WISE-optical color cut to further reduce stellar contamination. The quasar sample is almost perfectly covered in the sky by the Legacy photometric surveys as shown in Figure 1,  and we find the overlapped footprint has  an effective area of $4699\,\rm{deg}^{2}$.

\subsection{Completeness and Designs} \label{sec:obs:compl}
 As in \cite{2022ApJ...925...31X}, we employ the $z$-band $10\sigma$ point source depth to assess the mass completeness of photometric objects. This analysis is conducted within the matched footprint of the Legacy Surveys and the eBOSS quasar sample. As presented in Figure \ref{f1}, $90\%$ of the regions covered by the Legacy Surveys are deeper than $22.34$ in $z$ band. Therefore, we take $z=22.34$ as the galaxy depth  for the Legacy Surveys.   

\begin{figure}
	\centering
	\includegraphics[scale=0.55]{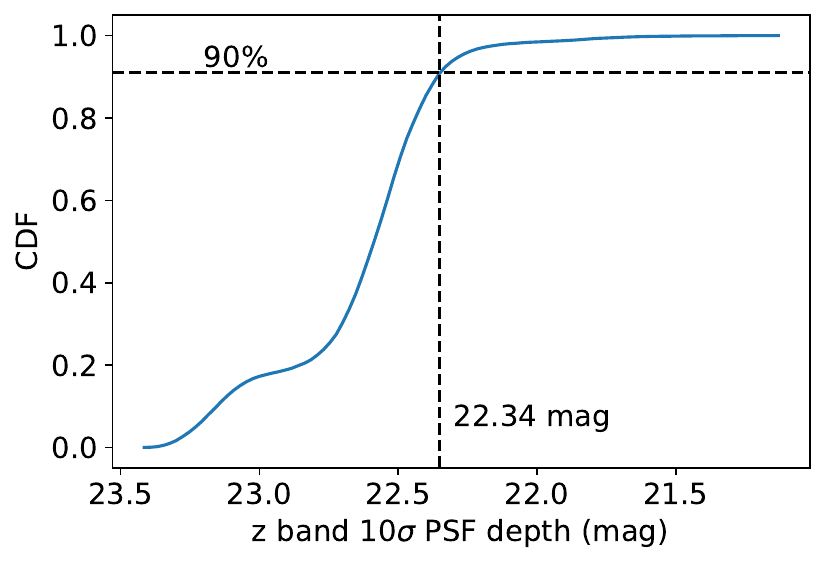}
	\caption{Cumulative distribution function of survey area to the $z$ band 10 $\sigma$ point source depth in the Legacy Surveys matched with eBOSS quasar footprint.}
	\label{f1}
\end{figure}

We calculate galaxy physical properties using the SED code {\texttt{CIGALE}} \citep{2019A&A...622A.103B} with $grz$ band fluxes. We adopt the stellar spectral library provided by \cite{2003MNRAS.344.1000B} to build up stellar population synthesis models. The initial stellar mass function (IMF) in  \cite{2003PASP..115..763C} is assumed. We take $Z/Z_{\odot} = 0.4, 1.0, 2.5$ as three different metallicities in our model. A delayed star formation history (SFH) $\phi(t) \simeq t \exp(-t/\tau)$ is taken, with the timescale $\tau$ varies from $10^7$ to $1.258 \times 10^{10}\,$yr with an equal logarithm space of 0.1$\,\rm{dex}$. We apply the starburst reddening \textbf{law} of \cite{2000ApJ...533..682C} to consider the dust attenuation, in which the color excess $E(B-V)$ changes from 0 to 0.5.

\begin{figure}
	\centering
	\includegraphics[scale=0.65]{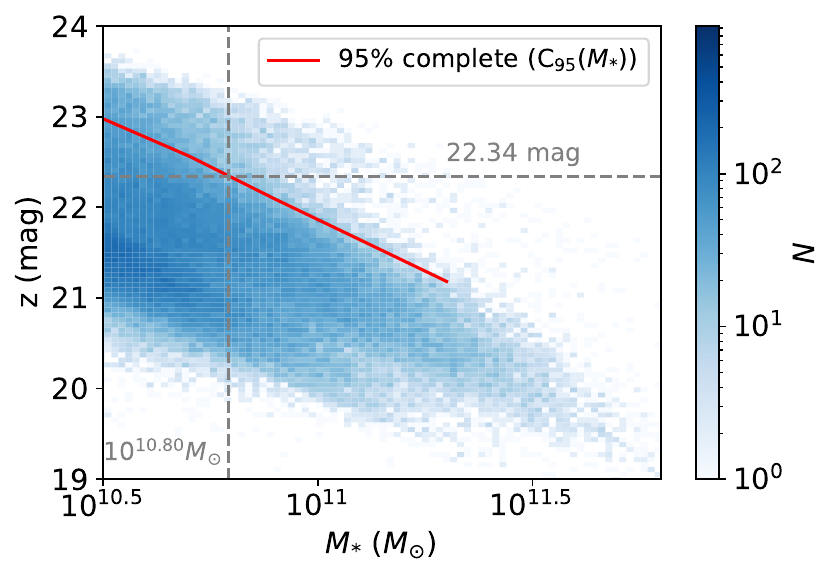}
	\caption{The distribution of stellar mass versus z-band magnitude for DECaLS DR9 galaxies with photo-z between 0.8 and 1.0. The red line represents the faint z band magnitude limit so that 95$\%$ of galaxies at each stellar mass are brighter than the limit. The horizontal gray dashed line shows the survey depth in z-band determined in Figure \ref{f1}. The vertical gray dashed line  is the completeness limit in stellar mass $10^{10.8} M_\odot$ for the photometric galaxies at the redshift of our interest.}
	\label{f2}
\end{figure}

 Following \cite{2022ApJ...925...31X}, we select the deepest $50\,\rm{deg}^2$ regions in the Legacy Surveys comprising 738 bricks, where the $z$-band 10$\sigma$ point source depth exceeds 23.37, to study the stellar mass completeness $C_{95}(M_{*})$ with photometric redshift ($z_{\rm{p}}$) calculated by \cite{2021MNRAS.501.3309Z}. $C_{95}(M_{*})$ is defined that $95\%$ of the galaxies are brighter than $C_{95}(M_{*})$ in $z$ band for a given stellar mass $M_{*}$. In Figure \ref{f2}, we show the stellar mass-$z$ band magnitude distribution of the deep photometric sample at redshift $0.8<z_{\rm{p}}<1.0$. We mark $C_{95}(M_{*})$ with red line in Figure \ref{f2}, from which the complete stellar mass is $10^{10.80}\,M_{\odot}$ at redshift $0.8<z_{\rm{s}}<1.0$ for the depth of $z$ band $22.34$. Hence, we use the stellar mass of photometric objects at the interval of $[10^{10.80},10^{11.80}]\,M_{\odot}$ at redshift $0.8<z_{\rm{s}}<1.0$ separated by four equal redshift bins. Furthermore, to constrain the SHMR at lower masses, we incorporate the incomplete stellar mass bins within the range of $[10^{10.00},10^{10.80}]\,M_{\odot}$ through appropriate modeling, as illustrated later.

\subsection{Measurements} \label{sec:obs:meas}
We follow the approach in \cite{2022ApJ...939..104X} to average the result at different sky regions and redshift bins with proper weights to acquire final $\bar{n}_2w_{\rm{p}}(r_{\rm{p}})$.

For representation simplicity, let $\mathcal{A}\equiv \bar{n}_2w_{\rm{p}}(r_{\rm{p}})$. Supposing $\mathcal{A}$ is calculated for $N_r$ redshift bins and $N_s$ sky regions, we further divide each region into $N_{\rm{sub}}$ sub-regions for error estimation. In our specific scenario, $N_r$ is set to 4, corresponding to the redshift bins [0.80,0.85], [0.85,0.90], [0.90,0.95] and [0.95,1.00]. Additionally, $N_s$ is defined as 3, representing BASS+MzLS NGC, DECaLS SGC and DECaLS NGC while  $N_{\rm{sub}}$ is established as 10.  According to Equation \ref{eaq:1}, $\mathcal{A}_{i,j,k}$ can be measured in the $i$th redshift bin, $j$th sky region and $k$th sub-sample through the Landy–Szalay estimator \citep{1993ApJ...412...64L}. Firstly, we average the measurements at different sky regions using the region area $w_{\rm{s},j}$ as a weighting factor to take the sky area difference into account:
\begin{equation}
     \mathcal{A}_{i,k}=\frac{\sum_{j=1}^{N_{\rm{s}}}\mathcal{A}_{i,j,k}w_{\rm{s},j}}{\sum_{j=1}^{N_{\rm{s}}}w_{\rm{s},j}}\,\,.
\end{equation}
Secondly, we obtain the mean values and the error vector $\sigma_{i}$ of the mean values for each redshift bin from $N_{\rm{sub}}$ sub-samples:
\begin{equation}
     \mathcal{A}_i=\sum_{k=1}^{N_{\rm{sub}}}\mathcal{A}_{i,k}/N_{\rm{sub}}\,\,,
\end{equation}
\begin{align}
    \sigma_{i}^{2} = \frac{1}{(N_{\rm{sub}}-1)N_{\rm{sub}}}\sum_{k=1}^{N_{\rm{sub}}}(\mathcal{A}_{i,k}(r_{\rm{p}})-\mathcal{A}_{i}(r_{\rm{p}}))^{2} \,\,,
\end{align}
Finally, results from different redshift bins are averaged
according to the error vector $\sigma_{i}$. Let $w_i(r_{\rm{p}})=\sigma^{-1}_i(r_{\rm{p}})$,
\begin{equation}
    \mathcal{A}=\frac{\sum_{i=1}^{N_{\rm{r}}}\mathcal{A}_{i}w_i^2}{\sum_{i=1}^{N_{\rm{r}}}w_i^2}\,\,,
\end{equation}
\begin{equation}
    \sigma=\sqrt{\frac{1}{\sum_{i=1}^{N_{\rm{r}}}{w_i^2}}}\,\,.
\end{equation}
We measure $\bar{n}_2w_{\rm{p}}(r_{\rm{p}})$ and the auto-correlation $w_{\rm{p}}$ of quasars in the radial range of $0.1\,h^{-1}{\rm{Mpc}}<r_{{\rm{p}}}<15\,h^{-1}\rm{Mpc}$. These scales are sufficiently broad to capture the distributions of both centrals and satellites. The measurements of $\bar{n}_2w_{\rm{p}}(r_{\rm{p}})$ are depicted as data points with error bars in Figure \ref{f3}. The error bars represent the vector $\sigma$. These measurements are overall good across all stellar mass bins within the entire radial range, except for the highest stellar mass bin $[10^{11.6},10^{11.8}] \, M_{\odot}$. Additionally, the auto-correlation $w_{\rm{p}}$ of quasars, with the weights assigned in the same way as  \citet{2021MNRAS.500.1201H}, is presented in Figure \ref{f31} to enhance the precision in constraining the mean host halo mass of quasars.

\begin{figure*}
	\centering
	\includegraphics[scale=0.60]{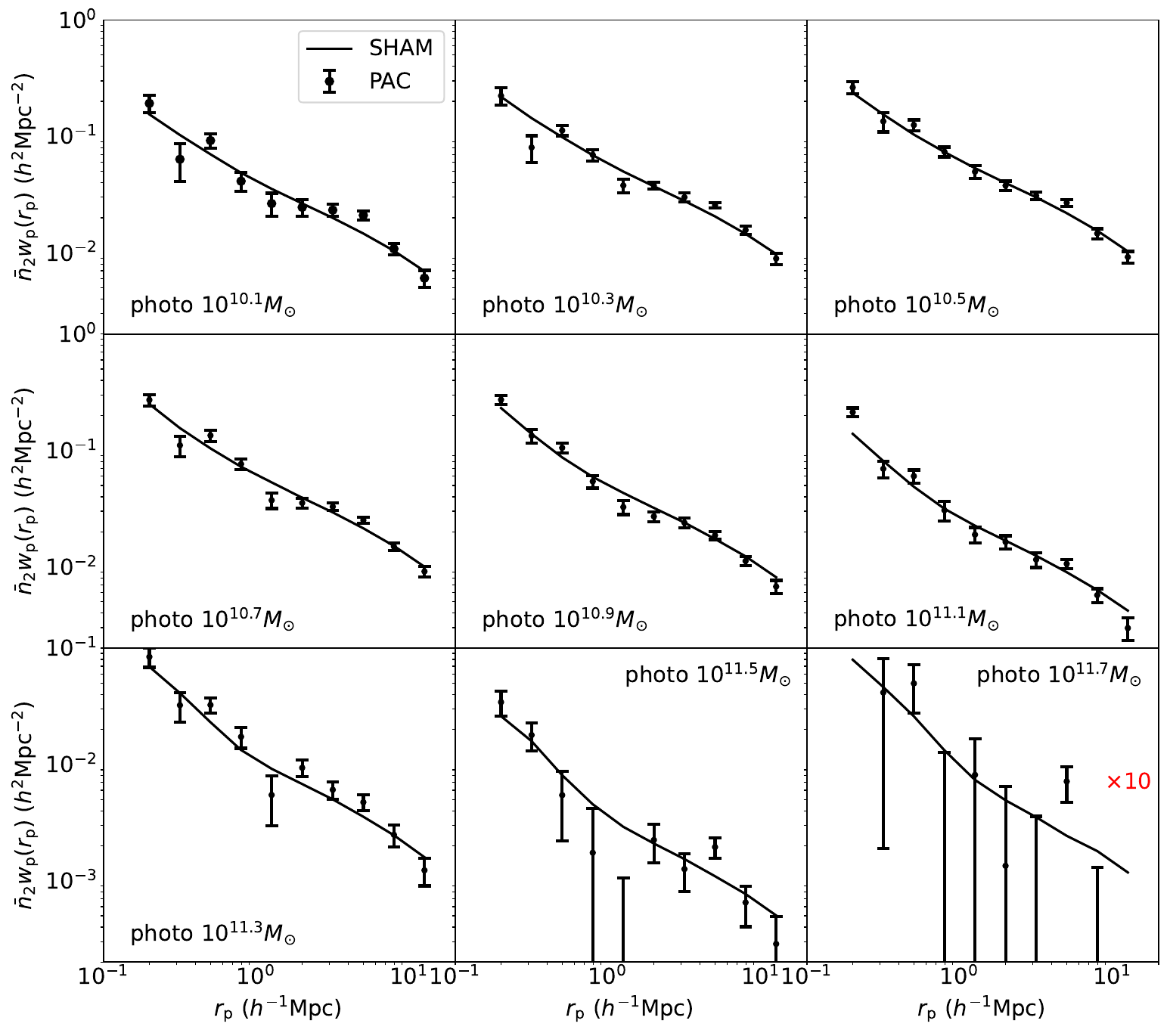}
	\caption{The measurement (data points with error bars) and the model fitting (lines) of the excess projected density $\bar{n}_2w_{\rm{p}}$ of galaxies around eBOSS quasars.  The stellar mass of galaxies ranges from $10^{10.0}\, M_{\odot}$ to $10^{11.8}\, M_{\odot}$ as the median mass indicated in each panel. The results for the highest stellar mass bin ($10^{11.7}\, M_{\odot}$) are multiplied by 10.} 
	\label{f3}
\end{figure*}

\begin{figure}
	\centering
	\includegraphics[scale=0.65]{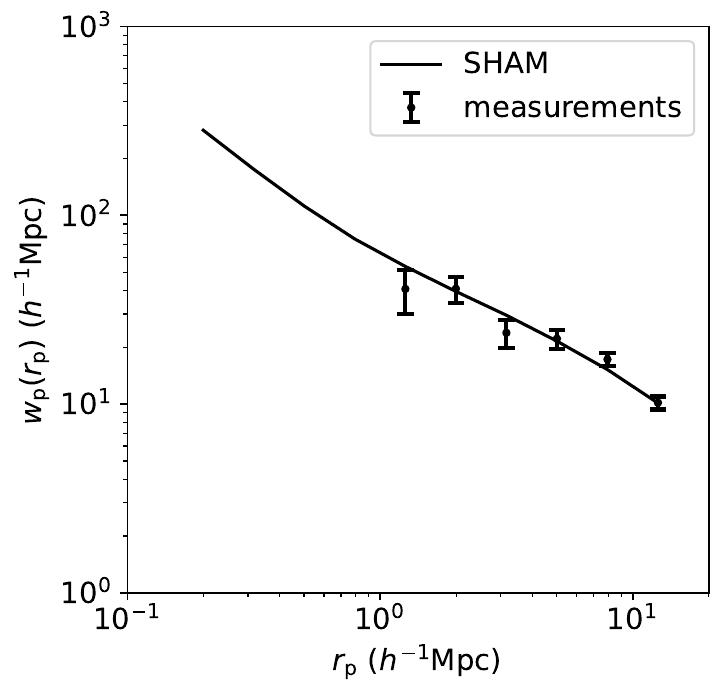}
	\caption{The measurement (data points with error bars) and the model fitting (line) of the projected auto-correlation $w_{\rm{p}}$ of quasars. } 
	\label{f31}
\end{figure}

\section{Simulation and Results} \label{sec:sim}

In this section, we outline the N-body simulation and the subhalo abundance matching (SHMR) method employed in this study for modeling the measurements. We present the best-fit results for the SHMR within the redshift range of $0.8<z_{\rm{s}}<1.0$ and explore the quasar-halo connection. Additionally, we highlight the high satellite fraction of quasars from our findings and assess the assumptions underlying the model.

\subsection{N-body Simulation} \label{sec:sim:nb}
We employ the \texttt{CosmicGrowth} simulation suite \citep{2019SCPMA..6219511J}, a grid of high-resolution N-body simulations utilizing the computationally-efficient adaptive parallel $\rm{P}^{3}\rm{M}$ method \citep{2002ApJ...574..538J,2021ApJ...915...75X}. We utilize the $\Lambda \mathrm{CDM}$ simulation with the following cosmological parameters: $\Omega_{\mathrm{m},0} = 0.268$, $\Omega_{\Lambda,0} = 0.732$, $h=0.71$, $n_{\mathrm{s}}=0.968$, and $\sigma_8=0.83$. This simulation comprises $3072^3$ dark matter particles within a $600 \, h^{-1} \mathrm{Mpc}$ box, with a softening length of $\eta = 0.01 \, h^{-1}\mathrm{Mpc}$. Groups are characterized with the friends-of-friends algorithm \citep{1985ApJ...292..371D}, employing a linking length of 0.2 times the mean particle separation. The halos are then processed with HBT+ \citep{2012MNRAS.427.2437H,2018MNRAS.474..604H} to identify subhalos and trace their evolution histories. Merger timescales of the subhalos with fewer than 20 particles are estimated using the fitting formula in \citet{2008ApJ...675.1095J}. Subhalos that have already merged into central subhalos are disregarded. The adopted simulation has a mass resolution of $m_{\mathrm{p}} = 5.54 \times 10^8 \,h^{-1}M_{\odot}$, which proves to be sufficiently fine for our study. We utilize snapshot 76 at a redshift of approximately 0.92 to align with the observations.

\subsection{Subhalo Abundance Matching} \label{sec:sim:sam}
To associate galaxies with (sub)halos in the N-body simulation, we implement the SHAM method. We utilize the widely-used five-parameter formula for the SHMR, a double power law with a constant scatter \citep{2010MNRAS.402.1796W,2012ApJ...752...41Y,2013MNRAS.428.3121M,2023ApJ...944..200X}:
\begin{equation}
    M_{*} = \left[\frac{2k}{(M_{{\rm{acc}}}/{M_0})^{-\alpha}+(M_{{\rm{acc}}}/{M_0})^{-\beta}}\right]\,.\label{eq:1}
\end{equation}
Here, $M_{{\rm{acc}}}$ is defined as the virial mass $M_{{\rm{vir}}}$ of the (sub)halo when it was last to become a central dominant object, where $M_{{\rm{vir}}}$ is defined through the fitting formula in \citet{1998ApJ...495...80B}. The scatter in $\log(M_*)$ at a given $M_{{\rm{acc}}}$ is assumed to follow a Gaussian distribution with a width of $\sigma$. The same set of parameters is applied for both centrals and satellites. \citep{2010MNRAS.402.1796W,2019MNRAS.488.3143B,2023ApJ...944..200X}. 

In addition to the standard SHAM parameters described above, given the inclusion of measurements from incomplete stellar mass bins, we introduce four additional parameters $k_{1}, k_{2}, k_{3}, k_{4}$ to account for the incompleteness of the four stellar mass bins. The modeled results from the simulation are obtained by multiplying the $\bar{n}_2w_{\rm{p}}(r_{\rm{p}})$ from the entire sample by the incompleteness, under the assumption that the incompleteness only affects the amplitude of the observed $\bar{n}_2w_{\rm{p}}(r_{\rm{p}})$.

To associate quasars with (sub)halos, we assume that the probability that a (sub)halo hosts a quasar follows a Gaussian distribution of logarithmic halo mass $\log_{10}[M_{{\rm{acc}}}/(h^{-1}M_{\odot})]$ with a mean value $\mu$ and a dispersion $\sigma_{\rm{q}}$. We use the same $\mu$ and $\sigma_{\rm{q}}$ for both halos and subhalos, Further we introduce an additional parameter $B$ to adjust the overall probability for a subhalo to host a quasar relative to a halo at the same mass.  This ensures that the subhalos have $B$ times probability to host a quasar relative to a halo of the same mass. Under the above schedule, the satellite fraction $f_{\rm{s}}$ of quasars is defined as:
\begin{equation}
    f_{\rm{sate}} = \frac{BN_{\rm{sate}}}{BN_{\rm{sate}} + N_{\rm{cen}}}\,\,. \label{eq:3}
\end{equation}
where $N_{\rm{sate}}$ and $N_{\rm{cen}}$ represent the numbers of subhalos and halos selected based on the same Gaussian distribution.

After assigning galaxies and quasars to (sub)halos, we compute the correlation functions using \texttt{Corrfunc} \citep{2020MNRAS.491.3022S} in the simulation. To compare with the measurements in observation, we define the $\chi^2$ as:
\begin{align}
    \chi^2&=\sum_{i=1}^{N_{\rm{m}}}\sum_{j=1}^{N_{\rm{r}}}(\frac{(\mathcal{A}^{{\rm{PAC}}}_{i}(r_{j})-\mathcal{A}^{{\rm{AM}}}_{i}(r_{j}))^2}{\sigma_{i}^2(r_{j})}\\ \notag
    &+ \sum_{j=1}^{N_{\rm{r}}}\frac{({w_{\rm{p}}}(r_{j})-{w_{\rm{p}}}^{{\rm{AM}}}(r_{j}))^2}{\sigma^2(r_{j})}\,\,.
\end{align}
 Here, $N_{{\rm_{m}}}$ and $N_{{\rm_{r}}}$ denote the numbers of stellar mass bins and of radial bins respectively. We explore the parameter space using the Markov Chain Monte Carlo (MCMC) sampler \texttt{emcee} \citep{2013PASP..125..306F}, employing maximum likelihood analyses for the three sets of parameters $\{M_{0},\alpha,\beta,k,\sigma\}$, $\{k_{1},k_{2},k_{3},k_{4}\}$, and $\{\mu, \sigma_{\mathrm{q}}, B\}$.

\subsection{SHMR and quasar-halo connection} \label{sec:sim:shmr}
The best-fit results and errors for the parameters are presented in the first row of Table \ref{t2}. Additionally, the joint posterior distributions of the parameters are illustrated in Figure \ref{f4} using \texttt{corner} \citep{2016JOSS....1...24F}. The best-fit $\bar{n}_2w_{\rm{p}}(r_{\rm{p}})$ and $w_{\rm{p}}$ results from the model are shown in Figure \ref{f3} and Figure \ref{f31} with solid lines. The fits are generally good for all stellar mass bins, both on the small and large scales. 

For quasars, we find that $\mu$, $\sigma_{\rm{q}}$, and $B$ are $13.01_{-0.13}^{+0.20}$, $0.65_{-0.12}^{+0.11}$, and $1.24_{-0.22}^{+0.26}$, respectively. The results indicate a very high satellite fraction of quasars, with $f_{\mathrm{sate}}=0.29_{-0.06}^{+0.05}$. $B$ is equal to 1 within the 1$\sigma$ error, implying that subhalos have nearly the same chance to host quasars as the halos, and the quasar activity is not affected by the larger halo environment. After determining the host halo masses for subhalos and adjusting the number density of quasars to match the observed value of approximately $1.29 \times 10^{-5} \ {h^{3}\mathrm{Mpc}^{-3}}$, we illustrate the HOD of our quasar sample in Figure \ref{f7}. We obtain median host halo masses for central and satellite quasars of $\mathrm{log}_{10}[M_{\rm{h},\rm{cen}}^{\rm{med}}/(h^{-1}M_{\odot})]=12.05_{-0.60}^{+0.60}$ and $\mathrm{log}_{10}[M_{\rm{h},\rm{sat}}^{\rm{med}}/(h^{-1}M_{\odot})]=12.90_{-0.72}^{+0.68}$. From the HOD, we observe that although the satellite fraction of quasars looks very high, each massive halo, on average, does not host more than one satellite quasar due to the low total number density of quasars. To more effectively depict the mass distributions, we also present the host halo mass distributions for central and satellite quasars in Figure \ref{f44}.

For galaxies, we present the SHMR at $0.8<z_{\rm{s}}<1.0$ in Figure \ref{f8} with $\mathrm{log}_{10}[M_{0}/(h^{-1}M_{\odot})]=11.83_{-0.07}^{+0.06}$, $\alpha=0.39_{-0.07}^{+0.05}$, $\beta=2.70_{-0.42}^{+0.17}$, $\mathrm{log}_{10}(k/M_{\odot})=10.21_{-0.06}^{+0.06}$ and $\sigma=0.25_{-0.05}^{+0.05}$, extending the findings from \citet{2023ApJ...944..200X} to higher redshifts. Our results underscore the potential of leveraging quasars as high-redshift tracers for investigating galaxy formation. However, larger quasar samples and deeper photometric catalogs are required to more precisely constrain the low mass end (e.g., $\beta$ in SHMR) and to explore higher redshifts.

Moreover, we determine the incompleteness of photometric samples for the four lowest stellar mass bins, yielding $k_{1}=0.43_{-0.05}^{+0.03}$, $k_{2}=0.62_{-0.05}^{+0.05}$, $k_{3}=0.70_{-0.05}^{+0.04}$ and $k_{4}=0.83_{-0.05}^{+0.04}$.  These results highlight the capability of the PAC method to constrain the stellar mass incompleteness of photometric samples and utilize the incomplete sample to explore lower mass regions.


\begin{figure}
	\centering
	\includegraphics[scale=0.67]{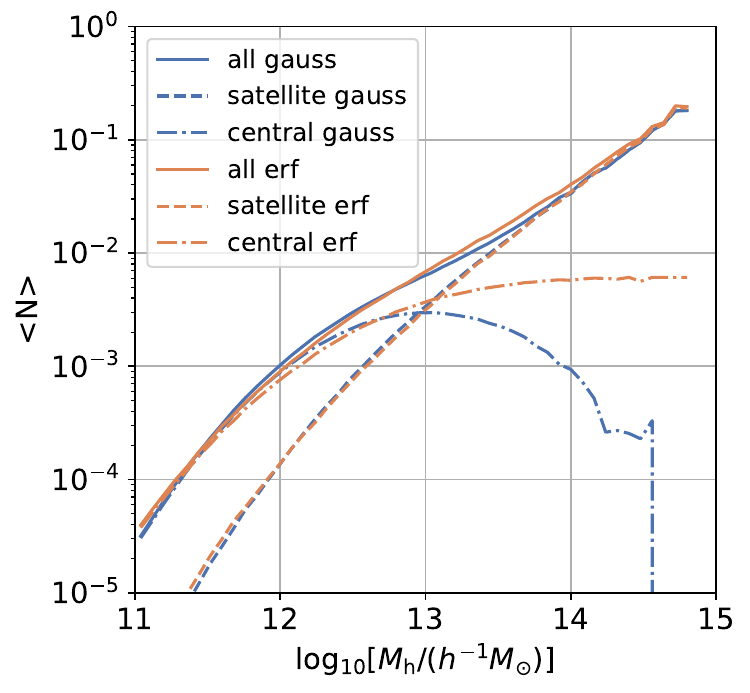}
	\caption{The mean HODs of quasars computed with the best-fit parameters. The results based on the Gaussian function for the quasar-halo relation are displayed in blue, with the parameters from the first row of Table \ref{t2}. For comparison, the results based on the error function are shown in orange, with the parameters from Table \ref{t3}. Central and satellite quasars are depicted by dotted and dot-dashed lines, respectively.} 
	\label{f7}
\end{figure}

\begin{deluxetable*}{cccccccccccc}
        \setlength{\tabcolsep}{0.7mm}
	\tablenum{1}
	\tablecaption{The best-fit parameter results and errors for the SHMR, incompleteness, and quasar-halo connection, assuming a Gaussian distribution for the quasar probability. The first row displays the fiducial results with all parameters free, while the second row presents the results for the remaining parameters after setting $\sigma_{\rm{q}}$ to 0.5. $M_{0}$ is in unit of $h^{-1}M_{\odot}$ and $k$ is in $M_{\odot}$. \label{t2}}
	\tablehead{ \colhead{$\mathrm{log}_{10}(M_{0})$} & \colhead{$\alpha$} & 
                    \colhead{$\beta$}&\colhead{$\mathrm{log}_{10}(k)$} & \colhead{$\sigma$} & \colhead{$\mu$} & \colhead{$\sigma_{\mathrm{q}}$}& \colhead{$k_{1}$} & \colhead{$k_{2}$} &
                    \colhead{$k_{3}$} &
                    \colhead{$k_{4}$}&
                    \colhead{$B$}
	}
	\startdata
        $11.83_{-0.07}^{+0.06}$ & $0.39_{-0.07}^{+0.05}$ & $2.70_{-0.42}^{+0.17}$& $10.21_{-0.06}^{+0.06}$ & $0.25_{-0.05}^{+0.05}$ & $13.01_{-0.13}^{+0.20}$&$0.65_{-0.12}^{+0.11}$& $0.43_{-0.05}^{+0.03}$& $0.62_{-0.05}^{+0.05}$ & $0.70_{-0.05}^{+0.04}$ & $0.83_{-0.05}^{+0.04}$& $1.24_{-0.22}^{+0.26}$\\
        $11.85_{-0.05}^{+0.05}$ & $0.38_{-0.05}^{+0.05}$ & $2.77_{-0.33}^{+0.10}$& $10.22_{-0.04}^{+0.05}$ & $0.25_{-0.04}^{+0.04}$ & $12.72_{-0.14}^{+0.12}$&$0.50 \ (\rm{set})$& $0.44_{-0.03}^{+0.02}$& $0.65_{-0.04}^{+0.03}$ & $0.71_{-0.04}^{+0.05}$ & $0.86_{-0.05}^{+0.05}$& $1.20_{-0.19}^{+0.24}$
        \enddata
\end{deluxetable*}

\begin{deluxetable*}{cccccccccccc}
        \setlength{\tabcolsep}{0.7mm}
	\tablenum{2}
	\tablecaption{The best-fit parameter results and errors for the SHMR, incompleteness, and quasar-halo connection, assuming an error function for the quasar probability. $M_{0}$ and $M_{\rm{min}}$ are in units of $h^{-1}M_{\odot}$ and $k$ is in $M_{\odot}$. \label{t3}}
 	\tablehead{ \colhead{$\mathrm{log}_{10}(M_{0})$} & \colhead{$\alpha$} & 
                \colhead{$\beta$}&\colhead{$\mathrm{log}_{10}(k)$} & \colhead{$\sigma$} & \colhead{$\mathrm{log}_
                {10}(M_{\mathrm{min}})$} & \colhead{$\sigma_{\mathrm{log}_{10}(M_{\rm{h}})}$}& \colhead{$k_{1}$} & \colhead{$k_{2}$} &
                    \colhead{$k_{3}$} &
                    \colhead{$k_{4}$}&
                    \colhead{$B$}
	}
	\startdata
        $11.97_{-0.07}^{+0.09}$ & $0.38_{-0.08}^{+0.06}$ & $2.23_{-0.32}^{+0.36}$& $10.34_{-0.08}^{+0.10}$ & $0.20_{-0.05}^{+0.04}$ & $12.81_{-0.44}^{+0.23}$&$1.01_{-0.16}^{+0.11}$& $0.45_{-0.05}^{+0.03}$& $0.71_{-0.06}^{+0.05}$ & $0.80_{-0.05}^{+0.06}$ & $0.91_{-0.05}^{+0.05}$& $1.22_{-0.18}^{+0.26}$        
	\enddata

\end{deluxetable*}

\subsection{Testing the robustness of the results} \label{sec:sim:robust}
The satellite fraction of quasars in our results is $f_{\mathrm{sate}}=0.29_{-0.06}^{+0.05}$. This high satellite fraction suggests that subhalos have nearly the same probability ($B=1.24_{-0.22}^{+0.26}$) as halos to host quasars for the same host (infall) mass, and the large-scale environment has little effect on quasar activity. To verify the robustness of such a high satellite fraction of quasars, we conducted three tests as follows. Since BASS is a bit shallower in g and r bands than the DECaLs, we measure $\bar{n}_2w_{\rm{p}}(r_{\rm{p}})$ in the DECaLS and BASS respectively, and a comparison of the results shows that the measurement is robust (Figure \ref{f4444}).

Initially, we observe a degeneracy between the parameters $\mu$ and $\sigma_{\rm{q}}$ in Figure \ref{f4}. To investigate the impact of this degeneracy on the satellite fraction of quasars, we fix $\sigma_{\rm{q}}$ at 0.50, a value consistent with \cite{2012ApJ...755...30R}, and constrain the remaining parameters through MCMC analysis. The resulting best-fit parameters are presented in the second row of Table \ref{t2}, and we find that a high satellite fraction is still necessary to align with the observations, and all parameters remain consistent with the previous results within the 1$\sigma$ interval. For better clarity, we depict the $\bar{n}_2w_{\rm{p}}$ results under these best-fit parameters for two stellar mass bins ($10^{10.5} \, M_{\odot}$ and $10^{10.9} \, M_{\odot}$) in Figure \ref{f32} using solid blue lines. We find that the measurements can also be well fitted by this model. This test confirms that the satellite fraction of quasars remains unaffected by the degeneracy between $\mu$ and $\sigma_{\rm{q}}$.

\begin{figure}
	\centering
	\includegraphics[scale=0.67]{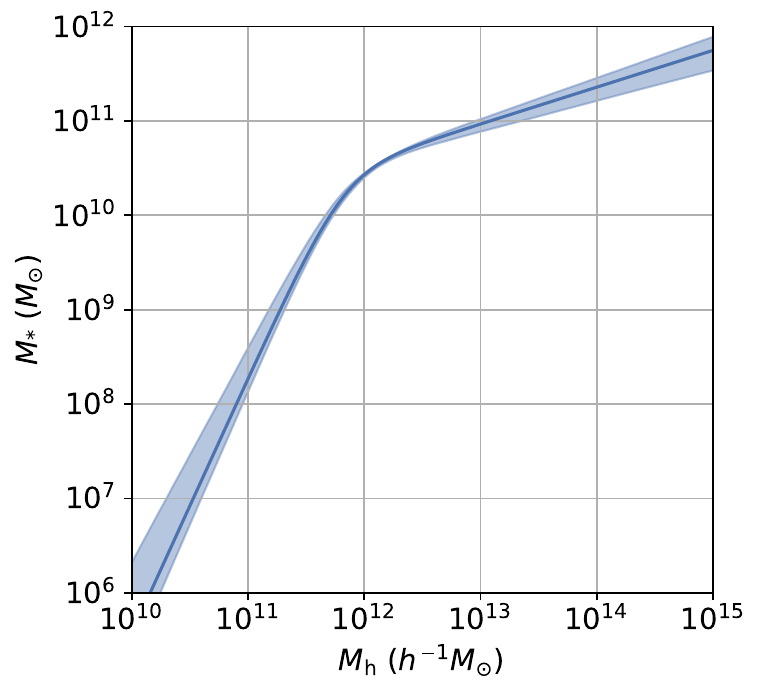}
	\caption{The stellar-halo mass relation at redshift $0.8<z_{\rm{s}}<1.0$. The mean SHMR and 1$\sigma$ errors are shown with blue line and shadow.} 
	\label{f8}
\end{figure}
\begin{figure*}
	\centering
	\includegraphics[scale=0.65]{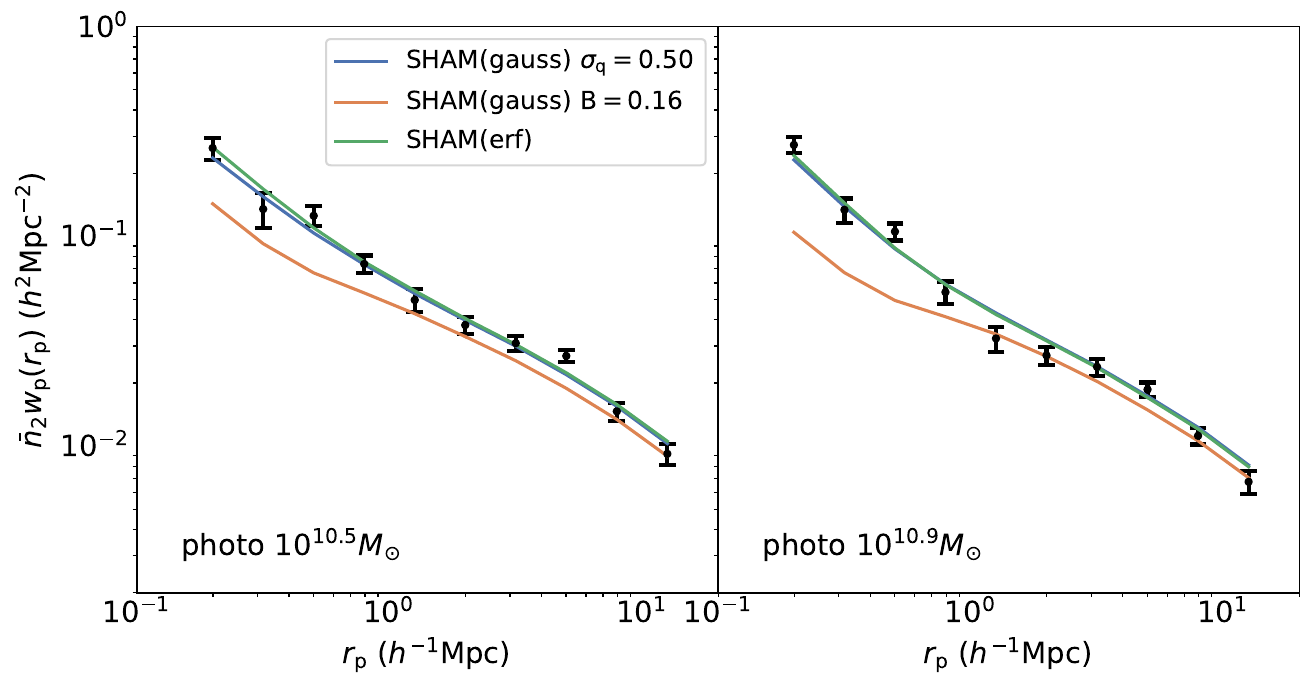}
	\caption{The $\bar{n}_{2}w_{\rm{p}}$ measurements in two stellar mass bins, $10^{10.5}\, M_{\odot}$ and $10^{10.9}\, M_{\odot}$. Dots with error bars represent the observational measurements. The orange solid lines depict the best-fit results with $f_{\rm{sate}}$ set to 0.05 (equivalently $B=0.16$), while keeping other parameters the same as in the first row of Table \ref{t2}. The blue solid lines show the best-fit results with $\sigma_{\rm{q}}$ set to 0.50. The green solid lines represent the best-fit results assuming a probability of a (sub)halo hosting a quasar modeled by an error function.} 
	\label{f32}
\end{figure*}
Secondly, to examine whether a low satellite fraction of quasars, as suggested by some previous studies \citep{2012ApJ...755...30R,2013ApJ...778...98S}, is consistent with our measurements, we set the $B$ parameter to be 0.16, corresponding to $f_{\mathrm{sate}}=0.05$ according to Equation \ref{eq:3}, while keeping the remaining parameters unchanged. The simulation results in two stellar mass bins ($10^{10.5}\, M_{\odot}$ and $10^{10.9}\, M_{\odot}$) are presented in Figure \ref{f32} with solid orange lines. In this case, we observe that the model fails to match the steep increase of $\bar{n}_2w_{{\rm{p}}}$ at small scales ($r_{\rm{p}}<1\,h^{-1}\rm{Mpc}$), suggesting that a higher satellite fraction is required to reproduce the observation.

Thirdly, considering our assumption of a Gaussian format for the quasar probability in our modeling, we aim to verify whether this assumption affects the satellite fraction of quasars. Therefore, we replace the Gaussian format with the error function format for both central and satellite quasars shown as follows:
\begin{equation}
     P(M_{\rm{h}})= \frac{1}{2}\left[1+ \mathrm{erf}(\frac{\mathrm{log}_{10}(M_{\rm{h}}) - \mathrm{log}_{10}(M_{\mathrm{min}})}{\sigma_{\mathrm{log}_{10}(M_{\rm{h}})}})\right]. \label{eq:4}
\end{equation}
where $P(M_{\rm{h}})$ is the probability of a halo becoming a quasar. $\mathrm{log}_{10}(M_{\mathrm{min}})$ and ${\sigma_{\mathrm{log}_{10}(M_{\rm{h}})}}$ are the characteristic mass scale and transition width of a softened step function. The best-fit parameters are presented in Table \ref{t3}, where we still find a high satellite fraction of quasars indicated by the parameter $B=1.22_{-0.18}^{+0.26}$ with $f_{\mathrm{sate}}=0.28_{-0.04}^{+0.05}$ . For a clearer illustration, simulation results in two stellar mass bins ($10^{10.5}\, M_{\odot}$ and $10^{10.9}\, M_{\odot}$) are shown in Figure \ref{f32} with solid green lines. We observe that the measurements can also be well-fitted by this model. To compare the difference between the Gaussian format and the error function, we show the HOD from the error function model in Figure \ref{f7}. We observe that HODs for satellites are nearly identical for these two models. Although some discrepancy is found for centrals with $M_{\rm{h}}>10^{13.0} \,M_{\odot}$, this does not significantly impact $f_{\rm{sate}}$ due to the low number densities of these massive halos. This test confirms that our results are not sensitive to the assumptions in the quasar-halo connection.

Our higher satellite fraction of quasars than that determined by \citet{2013ApJ...778...98S}  may come from two sources. One is the measurements of the cross correlation between galaxies and quasars. In this paper, we have used galaxies with stellar mass larger than  $10^{10.0}\, M_{\odot}$, compared with LRGs used by \citet{2013ApJ...778...98S}. While we find the slopes of the cross correlation functions within $r_{\rm{p}}=1\,h^{-1}\rm{Mpc}$ are consistent for massive galaxies (i.e. $>10^{11.2}\, M_{\odot}$) between the two works, our measurement for less massive galaxies also requires more satellite quasars around small central galaxies (cf. left panel of Figure \ref{f32}). The other source is different assumptions for the modeling. We use the abundance matching method, while \citet{2013ApJ...778...98S} used the HOD method. We note that the satellite fraction  is quite sensitive to the HOD forms used for the central quasars in \citet{2013ApJ...778...98S}. It changes from $f_{\rm sat}=0.068^{+0.034}_{-0.023}$ when the error function form is used  to  $f_{\rm sat}=0.099^{+0.046}_{-0.036}$ when the lognormal form is used instead. Furthermore they require that satellite quasars exist only in halos with $M_{\rm{h}}>10^{13.0} \,h^{-1}M_{\odot}$ (cf their Figure 14). In comparison, we allow many more small halos with $M_{\rm{h}}<10^{13.0} \,h^{-1}M_{\odot}$ (Figure \ref{f44}) to host satellite quasars which are required to reproduce the cross correlation between quasars and galaxies with stellar mass less than $10^{11.0}\, M_{\odot}$ (e.g. left panel of Figure \ref{f32}). Therefore, our higher satellite fraction mainly comes from those satellite quasars in halos with $M_{\rm{h}}<10^{13.0} \,h^{-1}M_{\odot}$ that were not probed by  \citet{2013ApJ...778...98S}.  Because central halos as small as $M_{\rm{h}}>10^{11.5} \,h^{-1}M_{\odot}$ can host quasars in the both works (cf. their Figure 14 and our Figure \ref{f7}), it is more reasonable that halos with $M_{\rm{h}}<10^{13.0} \,h^{-1}M_{\odot}$ can host satellite quasars, as there are subhalos that are massive enough.

\section{Conclusions} \label{sec:dis}

In this paper, we utilize a spectroscopic quasar sample from SDSS-IV eBOSS DR16 and a photometric galaxy sample from the Legacy Surveys. We employ the PAC method to measure $\bar{n}_{2}w_{\rm{p}}(r_{\rm{p}})$ between quasars and galaxies, in conjunction with the quasar auto-correlation $w_{\rm{p}}$, at redshift $0.8<z_{\rm{s}}<1.0$. Leveraging the advantages of PAC, we obtain reliable quasar clustering down to $0.1 \,h^{-1}\rm{Mpc}$. We model the measurements in N-body simulations using the SHAM approach and assume a Gaussian probability for a (sub)halo to host a quasar. We constrain the SHMR of galaxies and the quasar-halo connection. We verify the assumptions and confirm that a high satellite fraction of quasars is required to reproduce the observation. The main results are listed as follows:

\begin{enumerate}
	\item Under the assumption that the probability of a halo becoming a quasar follows a Gaussian distribution of logarithmic halo mass $\log_{10}[M_{{\rm{h}}}/(h^{-1}M_{\odot})]$, we find that the median host halo masses for central and satellite quasars are $\mathrm{log}_{10}[M_{\rm{h},\rm{cen}}^{\rm{med}}/(h^{-1}M_{\odot})]=12.05_{-0.60}^{+0.60}$ and $\mathrm{log}_{10}[M_{\rm{h},\rm{sat}}^{\rm{med}}/(h^{-1}M_{\odot})]=12.90_{-0.72}^{+0.68}$ at redshift $0.8<z_{\rm{s}}<1.0$ and a high satellite fraction of quasars of $f_{\mathrm{sate}}=0.29_{-0.06}^{+0.05}$. This high satellite fraction of quasars indicates that subhalos have nearly the same probability to host quasars as the halos for the same host (infall) mass, and the large-scale environment has little effect on the quasar activity. 
        \item We constrain the SHMR of galaxies at redshift $0.8<z_{\rm{s}}<1.0$, which can be described by a double power law with $\mathrm{log}_{10}[M_{0}/(h^{-1}M_{\odot})]=11.83_{-0.07}^{+0.06}$, $\alpha=0.39_{-0.07}^{+0.05}$, $\beta=2.70_{-0.42}^{+0.17}$, $\mathrm{log}_{10}(k/M_{\odot})=10.21_{-0.06}^{+0.06}$ and $\sigma=0.25_{-0.05}^{+0.05}$. These results underscore the potential of leveraging quasars as high redshift tracers for investigating galaxy formation.
\end{enumerate}

With the forthcoming data from DESI \citep{2023arXiv230606308D,2024AJ....167...62A} we anticipate gaining a better understanding of quasar clustering, given the larger number of surveyed quasars compared to SDSS-IV. Additionally, with the quasar-halo connection determined by our SHAM method, we can generate quasar mocks for DESI. With future large and deep photometric surveys like Legacy Survey of Space and Time (LSST) \citep{2019ApJ...873..111I} and Euclid \citep{2011arXiv1110.3193L}, we can gather much more quasar clustering information with smaller stellar mass bin.

\section*{Acknowledgments}
The work is supported by NSFC (12133006, 11890691), grant No. CMS-CSST-2021-A03, and 111 project No. B20019. We gratefully acknowledge the support of the Key Laboratory for Particle Physics, Astrophysics and Cosmology, Ministry of Education. This work made use of the Gravity Supercomputer at the Department of Astronomy, Shanghai Jiao Tong University.

Funding for the Sloan Digital Sky Survey IV has been
provided by the Alfred P. Sloan Foundation, the U.S.
Department of Energy Office of Science, and the Participating
Institutions. SDSS-IV acknowledges support and resources
from the Center for High-Performance Computing at the
University of Utah. The SDSS website is \url{www.sdss.org}.

SDSS-IV is managed by the Astrophysical Research
Consortium for the Participating Institutions of the SDSS
Collaboration including the Brazilian Participation Group, the
Carnegie Institution for Science, Carnegie Mellon University,
the Chilean Participation Group, the French Participation
Group, Harvard-Smithsonian Center for Astrophysics, Instituto
de Astrofísica de Canarias, The Johns Hopkins University,
Kavli Institute for the Physics and Mathematics of the Universe
(IPMU)/University of Tokyo, Korean Participation Group,
Lawrence Berkeley National Laboratory, Leibniz Institut für
Astrophysik Potsdam (AIP), Max-Planck-Institut für Astronomie (MPIA Heidelberg), Max-Planck-Institut für Astrophysik (MPA Garching), Max-Planck-Institut für Extraterrestrische
Physik (MPE), National Astronomical Observatories of China,
New Mexico State University, New York University, University of Notre Dame, Observatário Nacional/MCTI, The
Ohio State University, Pennsylvania State University, Shanghai
Astronomical Observatory, United Kingdom Participation
Group, Universidad Nacional Autónoma de México, University of Arizona, University of Colorado Boulder, University of
Oxford, University of Portsmouth, University of Utah,
University of Virginia, University of Washington, University
of Wisconsin, Vanderbilt University, and Yale University.

The Legacy Surveys consist of three individual and complementary projects: the Dark Energy Camera Legacy Survey (DECaLS; Proposal ID \#2014B-0404; PIs: David Schlegel and Arjun Dey), the Beijing-Arizona Sky Survey (BASS; NOAO Prop. ID \#2015A-0801; PIs: Zhou Xu and Xiaohui Fan), and the Mayall z-band Legacy Survey (MzLS; Prop. ID \#2016A-0453; PI: Arjun Dey). DECaLS, BASS and MzLS together include data obtained, respectively, at the Blanco telescope, Cerro Tololo Inter-American Observatory, NSF’s NOIRLab; the Bok telescope, Steward Observatory, University of Arizona; and the Mayall telescope, Kitt Peak National Observatory, NOIRLab. The Legacy Surveys project is honored to be permitted to conduct astronomical research on Iolkam Du’ag (Kitt Peak), a mountain with particular significance to the Tohono O’odham Nation.

\appendix
\restartappendixnumbering
\section{Posterior Distributions of the Parameters}\label{sec:A}
\begin{figure*}
	\centering
	\includegraphics[scale=0.27]{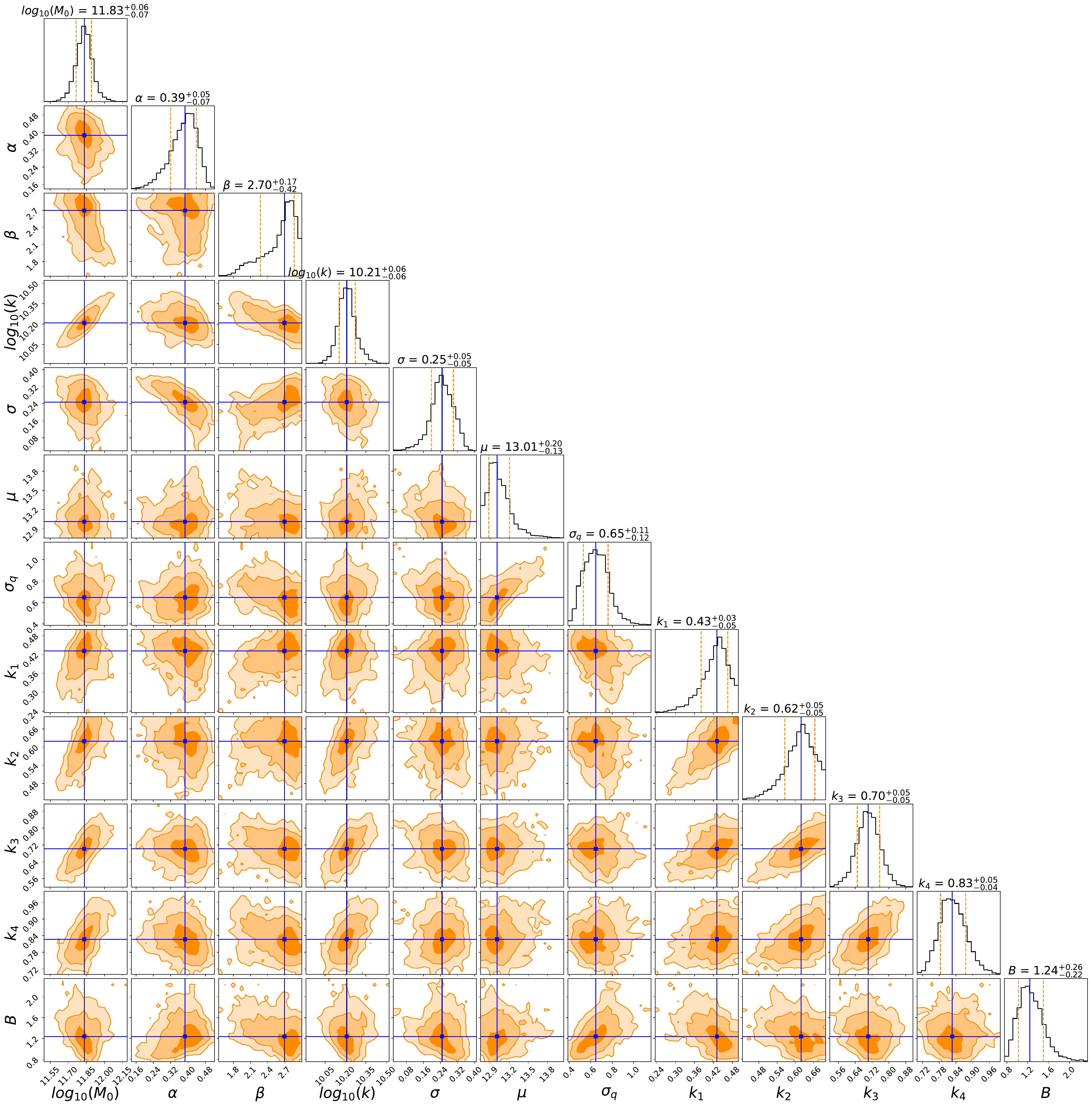}
	\caption{The posterior distributions of the 12 parameters in our model obtained through MCMC. The vertical green line indicates the median value of each parameter, and the dashed blue lines denote the 16th and 84th percentiles after marginalizing over the parameters.} 
	\label{f4}
\end{figure*}
We present the posterior PDFs of the 12 parameters in our fiducial model in this section, as shown in Figure \ref{f4}.
\restartappendixnumbering
\section{Host halo mass distributions of Quasars }\label{sec:B}
\begin{figure}
	\centering
	\includegraphics[scale=0.50]{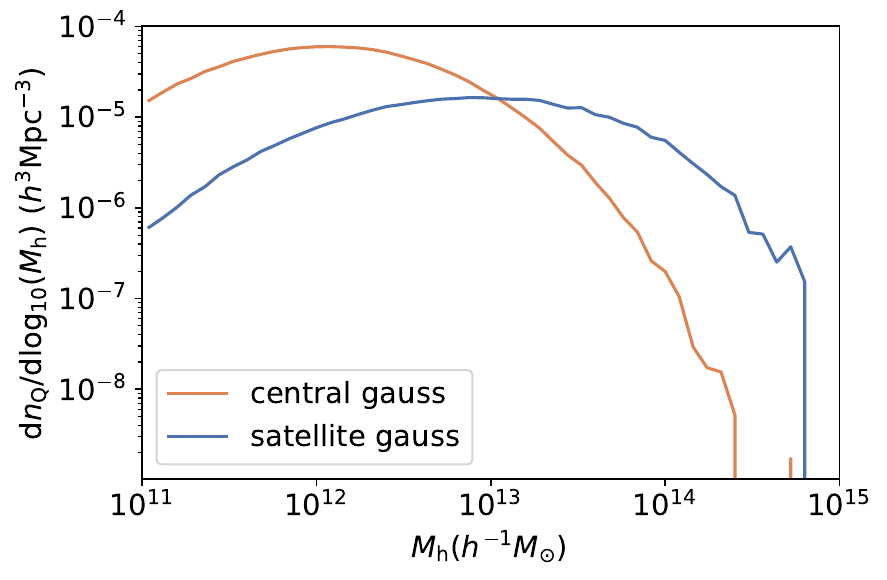}
	\caption{The number density distribution of host halos of central and satellite quasars. The central and satellite quasars are marked with orange and blue solid lines respectively.} 
	\label{f44}
\end{figure}
We present the mass distributions of host halos of central and satellite quasars in Figure \ref{f44}.
\restartappendixnumbering
\section{Comparison of $\bar{n}_{2}w_{\rm{p}}(r_{\rm{p}})$ measurements}\label{sec:C}
\begin{figure*}
	\centering
	\includegraphics[scale=0.57]{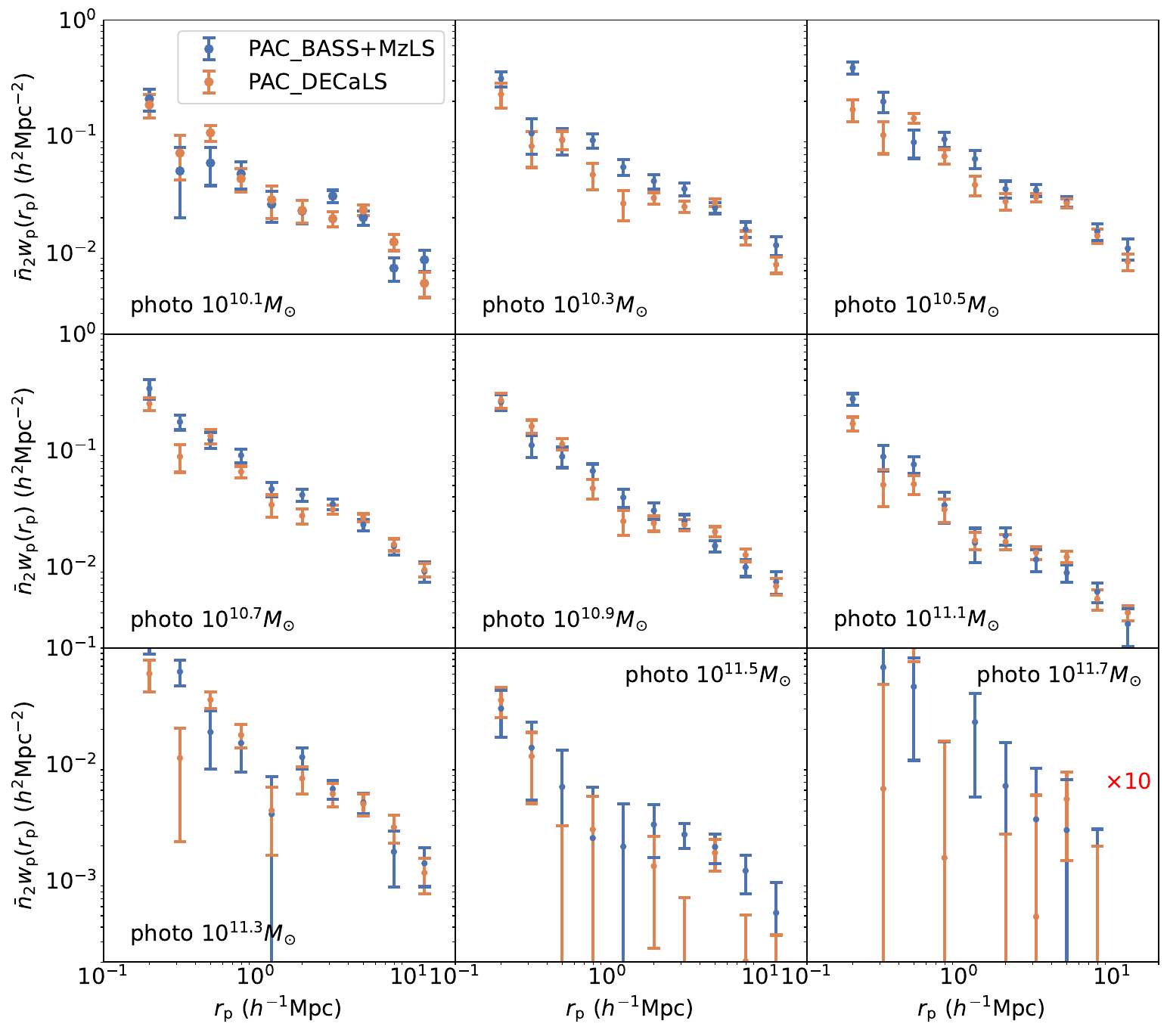}
	\caption{The measurement of $\bar{n}_{2}w_{\rm{p}}(r_{\rm{p}})$. Measurements with DR9 catalog from DECaLS are shown in blue colors and measurements with DR9 catalog from BASS and MzLS are shown in orange colors.} 
	\label{f4444}
\end{figure*}
We compare the $\bar{n}_{2}w_{\rm{p}}(r_{\rm{p}})$ measurements with  DR9 catalog from DECaLS and from BASS+MzLS that are shown in Figure \ref{f4444}.

\bibliography{sample631}{}
\bibliographystyle{aasjournal}


\end{document}